\documentclass{article}
\usepackage[a4paper=true,hypertex]{hyperref}
\usepackage{epsfig,amssymb,amsmath}
\usepackage{float}
\author{ Diego Harari, Silvia Mollerach, Esteban Roulet\\
{\it CONICET, Centro At\'omico Bariloche,}\\
{\it Av. Bustillo 9500 (8400) Argentina.}}

\title{Anisotropies of ultra-high energy cosmic ray nuclei \\
diffusing from extragalactic sources }

\begin{document}

\maketitle

\begin{abstract}
We obtain the dipolar anisotropies in the arrival directions of ultra-high energy cosmic ray nuclei diffusing from nearby extragalactic sources. We consider mixed-composition scenarios in which different cosmic ray nuclei are accelerated up to the same maximum rigidity, so that $E<ZE_\text{max}^p$, with $Z$ the atomic number and $E_\text{max}^p$  the maximum proton energy. We adopt  $E_\text{max}^p\simeq 6$ EeV so as to account for an increasingly heavier composition above the ankle. We obtain the anisotropies through Monte Carlo simulations that implement the cosmic ray diffusion in extragalactic turbulent fields as well as the effects of photo-disintegrations and other energy losses.  Dipolar anisotropies at the level of 5 to 10\% 
at energies $\sim 10$~EeV are predicted for plausible values of the source density and magnetic fields.

\end{abstract}

\section{Introduction}

In a previous work, hereafter referred to as paper I \cite{I}, we studied the expected dipolar anisotropies  of ultra-high energy cosmic rays (UHECRs) diffusing from nearby extragalactic sources, under the assumption that the cosmic rays were  protons.   In this work we extend this analysis to the case  of heavier nuclei, which is of relevance due to the indications provided by measurements with the Pierre Auger Observatory of a transition towards a heavier composition at energies larger than a few EeV \cite{augercomposition1,augercomposition2} (where 1 ${\rm EeV\equiv 10^{18}}$ eV). 

We will consider scenarios in which a mixed composition is present at the sources and such that the acceleration of the different nuclei occurs up to the same maximum rigidity, as  expected if the acceleration process is of electromagnetic nature.  Assuming for simplicity a sharp cutoff,
 the cosmic ray (CR) energies satisfy in this case  $E<ZE_\text{max}^p$,   where $Z$ is the atomic number and $E_\text{max}^p$is the maximum proton energy adopted. This would then naturally give rise to a transition towards heavier elements for energies $E>E_\text{max}^p$. We will adopt hereafter $E_\text{max}^p\simeq 6$~EeV, the so-called low rigidity scenario \cite{Aloisio2011,Allard2011,Mollerach2013}, for which the transition to heavier elements happens just above the ankle of the spectrum (which is the observed transition towards a harder spectrum taking place at an energy $\sim 5$~EeV). In this case heavy nuclei such as Fe, which have $Z=26$, can reach energies  of order 150 EeV, comparable to the highest CR energies observed. Note that given the relatively small maximum rigidities considered, the conditions required for the properties of the sources are less extreme in these scenarios than in the ones in which protons are accelerated up to beyond 100 EeV. 

An important observable which may help to discriminate between different scenarios is the anisotropy in the arrival direction distribution at different angular scales.  In particular, the low level of anisotropies observed at small angular scales at the highest energies \cite{Augerdensitybounds,AugerAnisoHE} ($E>40$ EeV) is somewhat unexpected in proton scenarios,  for which the deflections in galactic and extragalactic magnetic fields would be of only a few degrees. Some hints of intermediate angular scale excesses appearing at the highest energies have been reported by the Telescope Array \cite{TAanisoHE} (a hot spot on a 20$^\circ$ radius circle for $E>57$~EeV) and by the Auger Observatory \cite{AugerAnisoHE} (an excess on a 12$^\circ$ radius circle close to the Super-Galactic Plane and centered at about 18$^\circ$ from
the direction of Centaurus A, for $E > 54$~EeV). Finally, searches for anisotropies on large angular scales, which look for the presence of a dipole or  a quadrupole, are of interest since they could arise  in cases in which deflections are large and cosmic rays diffuse, and eventually also in the case of quasi-rectilinear propagation due to the anisotropic distribution of the nearby sources. The latest results from the Pierre Auger Observatory indicate some hints of a $\sim 7$\% dipolar anisotropy for energies $E>8$ EeV \cite{AugerDipole}.

\section{Effects of background radiations upon UHECR nuclei}

One of the difficulties that appear when trying to obtain predictions in scenarios involving UHECR nuclei is that during their propagation they interact with the radiation backgrounds, both the cosmic microwave background (CMB) as well as the extragalactic background light (EBL) which contributes mostly at infrared and visible wavelengths. As a result, nuclei can photo-disintegrate and  change their mass. In the case of protons one just needs to solve an equation describing the energy losses, what leads to changes in the Lorentz factor $\Gamma$ due to redshift effects as well as the photo-pion and pair creation losses. Instead, in the case of nuclei one also needs to consider the change in the nuclear mass by the stochastic effects of photo-disintegrations. 

Since we will consider scenarios with relatively small maximum energies, $E<6Z$~EeV, and we will focus on CR energies near and above the ankle, $E>4$ EeV,
then the relevant Lorentz factors for this work will be $7\times 10^{7}<\Gamma<6\times 10^{9}$. In particular, this implies that photo-pion production processes will be negligible, since for interactions with the CMB the pion production threshold would require values $\Gamma>4\times 10^{10}$,  while the photo-pion production is negligible in the case of interactions with the EBL.

For the photo-disintegration processes we will adopt the cross section parameterizations given by Puget, Stecker and Bredekamp (PSB) \cite{Puget1976}, with the improved threshold energies reported in \cite{Stecker1999}. Given the maximum energies considered, $E<6Z$ EeV, and the fact that we will only be interested in energies  $E>4$ EeV, we can also safely neglect the secondary nucleons emitted in the photo-disintegration process, since they will have energies smaller than $E_\text{max}^p Z/A\simeq 3$ EeV. This simplifies the analyses since we will just need to follow the leading fragment after the photo-disintegration. We will also consider that  for any given nuclear mass we can assume that the relevant nucleus is the one corresponding to  the most stable isotope.

The rate at which a nucleus of mass number $A$ photo-disintegrates with the emission of $i$ nucleons is
\begin{equation}
R_{A,i}=\frac{1}{2\Gamma^2}\int_0^\infty \frac{\text{d}\varepsilon}{\varepsilon^2}\frac{\text{d}n_\gamma}{\text{d}\varepsilon}\int_0^{2\Gamma\varepsilon}{\text{d}}\varepsilon' \,\varepsilon'\,\sigma_{A,i}(\varepsilon'),
\end{equation}
where the cross sections $\sigma_{A,i}$ were parameterized in \cite{Puget1976}. The photon number density d$n_\gamma/\text{d}\varepsilon$ includes both the CMB and the EBL contributions. An important ingredient in these studies is the modelling of the EBL and its redshift evolution, for which we will adopt the values reported in the work of Inoue et al. \cite{Inoue2013}.

To describe the average mass loss of a given nucleus it is convenient to introduce the effective rate
\begin{equation}
R_{A,eff}\equiv \sum_i i R_{A,i}.
\end{equation}
In terms of this rate one has that the average change in the mass number after traversing a distance d$x$ is
\begin{equation}
{\rm d}A=-R_{A,eff}\,{\rm d}x.
\end{equation}
The attenuation length with respect to the photo-disintegrations can be defined as
\begin{equation}
\lambda_{pd}\equiv\left[-\frac{1}{A}\frac{\text{d}A}{\text{d}x}\right]^{-1}=\frac{A}{R_{A,eff}}.
\end{equation}
We plot in Fig. 1 this attenuation length for different nuclei (He, C, Si and Fe) and for two values of the redshift, $z=0$ and 0.5, as a function of the Lorentz factor $\Gamma$. Also shown in the plot is the attenuation length for pair creation processes, $\lambda_{pair}$, computed using the fit obtained in paper I for the case of protons and using the scaling relation $\lambda_{pair}^A(\Gamma)\simeq A\lambda_{pair}^p(\Gamma)/Z^2$, which holds up to small effects associated to Coulomb corrections.
\begin{figure}[t]
\centerline{\epsfig{width=3.in,angle=-0,file=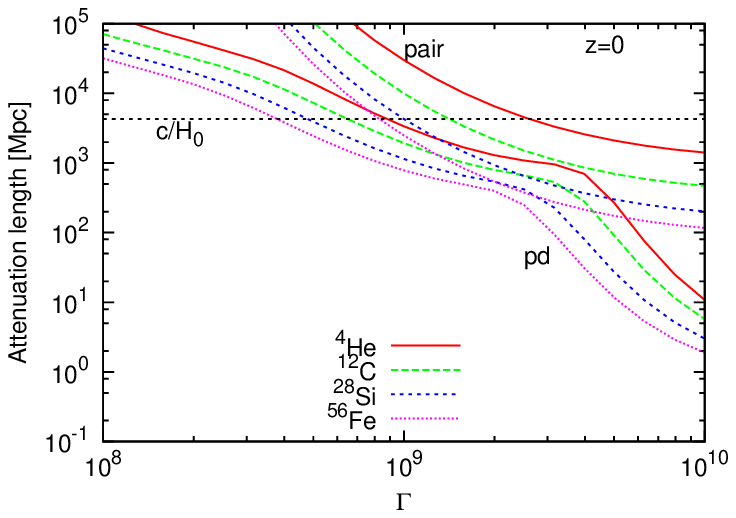}\epsfig{width=3.in,angle=-0,file=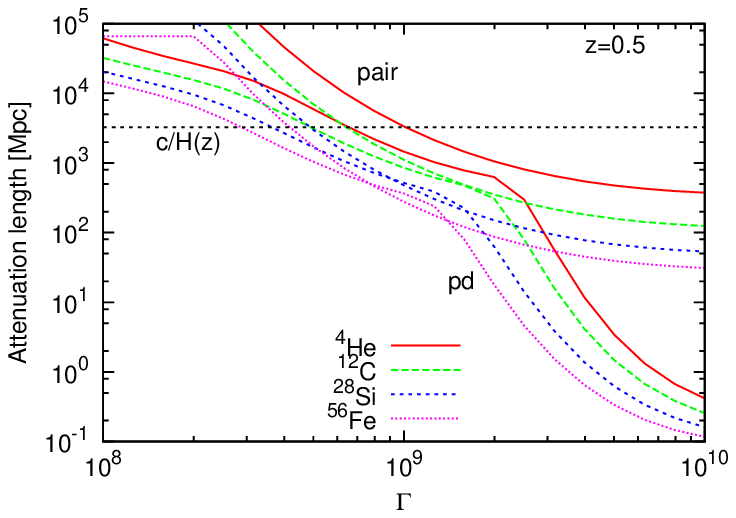}}
\vskip 1.0 truecm
\caption{Attenuation lengths for different nuclei (He, C, Si and Fe) and for two values of the redshift, $z=0$ (left) and $z=0.5$ (right), as a function of the Lorentz factor $\Gamma$. } 
\label{r2t}
\end{figure}

Using that $E\simeq \Gamma A m_p$, with $m_p$ the proton mass,  and computing the evolution as a function of redshift so as to account more directly for cosmological effects, one has that
\begin{equation}
\frac{1}{E}\frac{{\rm d}E}{{\rm d}z}=\frac{1}{\Gamma}\frac{{\rm d}\Gamma}{{\rm d}z}+\frac{1}{A}\frac{{\rm d}A}{{\rm d}z},
\end{equation}
where
\begin{equation}
\frac{1}{\Gamma}\frac{{\rm d}\Gamma}{{\rm d}z}=-c\frac{{\rm d}t}{{\rm d}z}
\left[\frac{1}{\lambda_{pair}}+\frac{1}{\lambda_{ad}}\right],
\end{equation}
and
\begin{equation}
\frac{1}{A}\frac{{\rm d}A}{{\rm d}z}=-c\frac{{\rm d}t}{{\rm d}z}\frac{1}{\lambda_{pd}}.
\end{equation}
In the previous expressions
\begin{equation}
\frac{{\rm d}t}{{\rm d}z}=-\frac{1}{(1+z)H(z)},
\end{equation}
with the Hubble constant being
\begin{equation}
H(z)=H_0\sqrt{(1+z)^3\Omega_m+\Omega_\Lambda}.
\end{equation}
We will adopt present values $H_0=70~{\rm km\, s^{-1}\, Mpc^{-1}}$ for the Hubble constant, $\Omega_m=0.3$ for the matter contribution and $\Omega_\Lambda=0.7$ for the vacuum energy contribution. The attenuation length for the adiabatic losses is $\lambda_{ad}=c/H(z)$. Note that the attenuation lengths introduced above have the meaning of an energy loss length, i.e. represent the distance over which the energy would change to $1/e$ of its initial value if the attenuation lengths were to remain constant. On the other hand, the typical length for the emission of one nucleon by photo-disintegration is $\lambda_{pd}/A$.

In order to account for the stochastic nature of the photo-disintegration processes while following the propagation of a given nucleus, we will assume that the probability of emitting one nucleon\footnote{In the particular case of $^9$Be we account for the fact that the photo-disintegration produces two alpha particles and a nucleon, and for simplicity we assume that the two alpha particles follow similar trajectories afterwards.} in a given step d$x$ is just $[1-\exp(-{A\rm d}x/\lambda_{pd})]$. In this way we will consider that individual nucleons are emitted stochastically with the effective rate that already accounts also for the probability of multi-nucleon emission.

\section{Diffusion of UHECR nuclei in turbulent magnetic fields}

At the energies considered in this work, $E>4$~EeV, CRs are most likely of extragalactic origin, and our aim here is to study the effects that could arise if they diffuse from their sources in the presence of sizable turbulent extragalactic magnetic fields present in the Local Supercluster. Extragalactic magnetic fields are expected to be produced  either by the evolution  of primordial seeds during the process of structure formation or as a result of galactic outflows. Their strength should then be enhanced in denser regions while suppressed in the voids. A turbulent component in the Local Supercluster with RMS values of 1--10 nG, with typical coherence lengths $l_c\sim 1$~Mpc, is often considered, and we will here adopt the illustrative values of $B=1$~nG and $l_c=1$~Mpc, taken to be  homogeneous over space. In Section 7 we will analyze the dependence of the anisotropy on the assumed strength of the magnetic field. Note that since CRs will be diffusing in this scenario, the most relevant magnetic field effects will be those affecting relatively nearby sources, i.e. within $\sim 200$~Mpc. Contributions from sources farther away will be suppressed either by a magnetic horizon effect at low energies or by the interactions with the radiation backgrounds at the highest energies. For definiteness we will consider that the turbulent field has a Kolmogorov spectrum, so that the magnetic field energy density scales as $\omega(k)\propto k^{-5/3}$ in Fourier space. An important quantity (see paper I for details) to characterize the diffusion effects is the critical energy $E_c$ corresponding to the value for which the Larmor radius $r_L$ becomes equal to $l_c$, given by
\begin{equation}
E_c= ZeBl_c\simeq 0.9Z\frac{B}{\rm nG}\frac{l_c}{\rm Mpc}\,\text{EeV}.
\end{equation}

Particles with $E<E_c$ undergo large deflections caused by the magnetic field modes with scales comparable to the Larmor radius, giving rise to the so-called resonant diffusion. Instead, particles with $E>E_c$ are in the regime in which the deflections over each coherence length, $\delta\simeq l_c/r_L$,  are small. Hence,  only after the CRs traverse several coherence domains, over a distance scale known as the diffusion length $l_D$, the overall deflection becomes $\sim 1$~rad. At distances larger than $l_D\equiv 3D/c$ spatial diffusion  takes place, characterized by the (isotropic) diffusion coefficient $D$. In the regime with $E>E_c$ there is then non-resonant diffusion if the source distance $r_s$ is such that $r_s\gg l_D$,  while if $r_s<l_D$, which is the case at sufficiently high energy, the propagation is quasi-rectilinear.

For the parameters considered in this work we have  $E_c\simeq 0.9 Z$~EeV at present and hence for $E>4$~EeV only the heavy nuclei may be in the resonant regime. Note that the coherence length typically has a redshift dependence $l_c(z)=l_c(0)/(1+z)$ while the magnetic field is generally enhanced for higher redshifts as $B(z)\simeq B(0)(1+z)^{2-\mu}$, where the parameter $\mu$ was introduced in \cite{be07} to account for MHD effects. Following \cite{be07} we will assume $\mu=1$, in which case the critical energy actually results independent of redshift.

The observed distribution of arrival directions of nuclei that originate from an individual source depends on the magnetic field they traverse, the energy of the particles and  the distance to the source.  In the diffusion regime the dipolar anisotropy is given by $\vec \Delta = 3 D \vec \nabla n / n$, where $n$ is the spatial density of particles. In paper I we used this expression to compute $\vec\Delta$, with $n$ the solution of the diffusion equation for protons, including energy losses during propagation, obtained following Berezinsky and Gazizov \cite{be06,be07}. We also showed that the dipolar anisotropy can alternatively be computed through numerical simulations of the trajectories of particles propagating in a turbulent magnetic field in terms of the average of the cosine of the angle $\theta$ between the 
final CR velocity and the vector describing the particle position with respect to the source. The dipole amplitude is $\Delta=3\langle \cos \theta \rangle$. This method has the advantage that it can be used for all energies, encompassing the transition from the diffusive propagation of low rigidity particles, that induces small amplitude anisotropies, to the quasi-rectilinear regime at high rigidities, that leads to large anisotropies. We implemented this method through a numerical integration of the stochastic differential equation that describes the scattering of UHECRs in a turbulent magnetic field \cite{ac99,oksendal,gardiner}:
\begin{equation}
{\rm d}n_i =-\frac{1}{l_D}n_i c{\rm d}t + \frac{1}{\sqrt{l_D}} P_{ij}{\rm  d}W_j,
\label{dni}
\end{equation}
where $P_{ij} \equiv (\delta_{ij} - n_i n_j)$ is the projection tensor onto the plane orthogonal to $\hat n\equiv (n_1,n_2,n_3)$ (the direction of the CR velocity),
repeated indices are summed, and (${\rm d}W_1,\,{\rm d} W_2,\, {\rm d}W_3$) are three Wiener processes such that $\langle{\rm d} W_i \rangle =0$ and  $\langle {\rm d}W_i{\rm d}W_j \rangle =c{\rm d}t\delta_{ij}$.

In the present paper we use the stochastic method across the entire range of rigidities considered to obtain $\cos\theta$ for each of the propagated particles. We use as input for $l_D=3D/c$ the diffusion coefficient $D(E)$ evaluated in paper I through numerical integration of the Lorentz equation for the trajectories of charged particles in a homogeneous turbulent magnetic field. The analytic fit to the results for the case of Kolmogorov turbulence (that were comparable to those in ref.~\cite{gl07}) is given by

\begin{equation}
D(E) = \frac{c}{3}l_c\left[ 4\left(\frac{E}{E_c}\right)^2 + 0.9\left(\frac{E}{E_c}\right) + 0.23\left(\frac{E}{E_c}\right)^{1/3}\right] \ .
\label{D(E)}
\end{equation}

If energy losses and cosmological expansion effects are neglected, the dipole anisotropy evaluated as the average  $\Delta=3\langle \cos \theta \rangle$  over a large set of trajectories obtained with a numerical integration of equation (\ref{dni}) coincides with $\Delta=3 D(E)/cr_s$ for  $r_s > l_D$, as expected in the diffusion regime. In the case of $r_s$ much smaller than $l_D$ it coincides with the approximation 
$\Delta/3 \simeq 1-c{r_s}/({9 D(E)})$ valid for the quasi-rectilinear regime (an approximate expression for the anisotropy valid for all source distances and energies is given by eq. (21) in paper I). The use of this method allows to evaluate the anisotropy across the different regimes, accounting also for energy losses and nuclear fragmentation, as we do in the next sections. 

\section{Results for individual sources and for each type of nuclei}

We first compute the dipolar anisotropy from an individual source that injects  just one type of  nucleus, as a function of the source distance and the arrival energy. To cover the relevant range of possible masses we consider five different representative nuclei: p, He, C, Si and Fe. In each case we propagate particles with energies ranging from  $E_{min}= 4$ EeV up to  $E_{max}= 6Z$ EeV, with a spectral distribution $E^{-2}$ and a sharp cutoff at $E_{max}$.  We inject at the source a constant number of particles per unit of time, neglecting for simplicity source evolution effects, and we integrate the contributions from a maximum redshift $z_{max}=1$ down to  the present, $z=0$ (note that in the diffusion regime the redshift has the meaning of time rather than distance). We follow the evolution of the mass number (and charge) and of the Lorentz factor of the particle as a function of the redshift as described in Section 2. We also follow the spatial trajectory as described in Section 3, solving equation (\ref{dni}) numerically, and evaluating $\cos\theta$ for each propagated particle.  When the particle arrives to $z=0$ we record its energy, mass, distance from the source and the angle between the  source direction and the CR arrival direction. We can then obtain the dipolar amplitude of nuclei by computing $\langle \cos\theta\rangle$ averaged over the set of particles with specified arrival energy and source distance.

\begin{figure}[t]
\centerline{\epsfig{width=2.2in,angle=-90,file=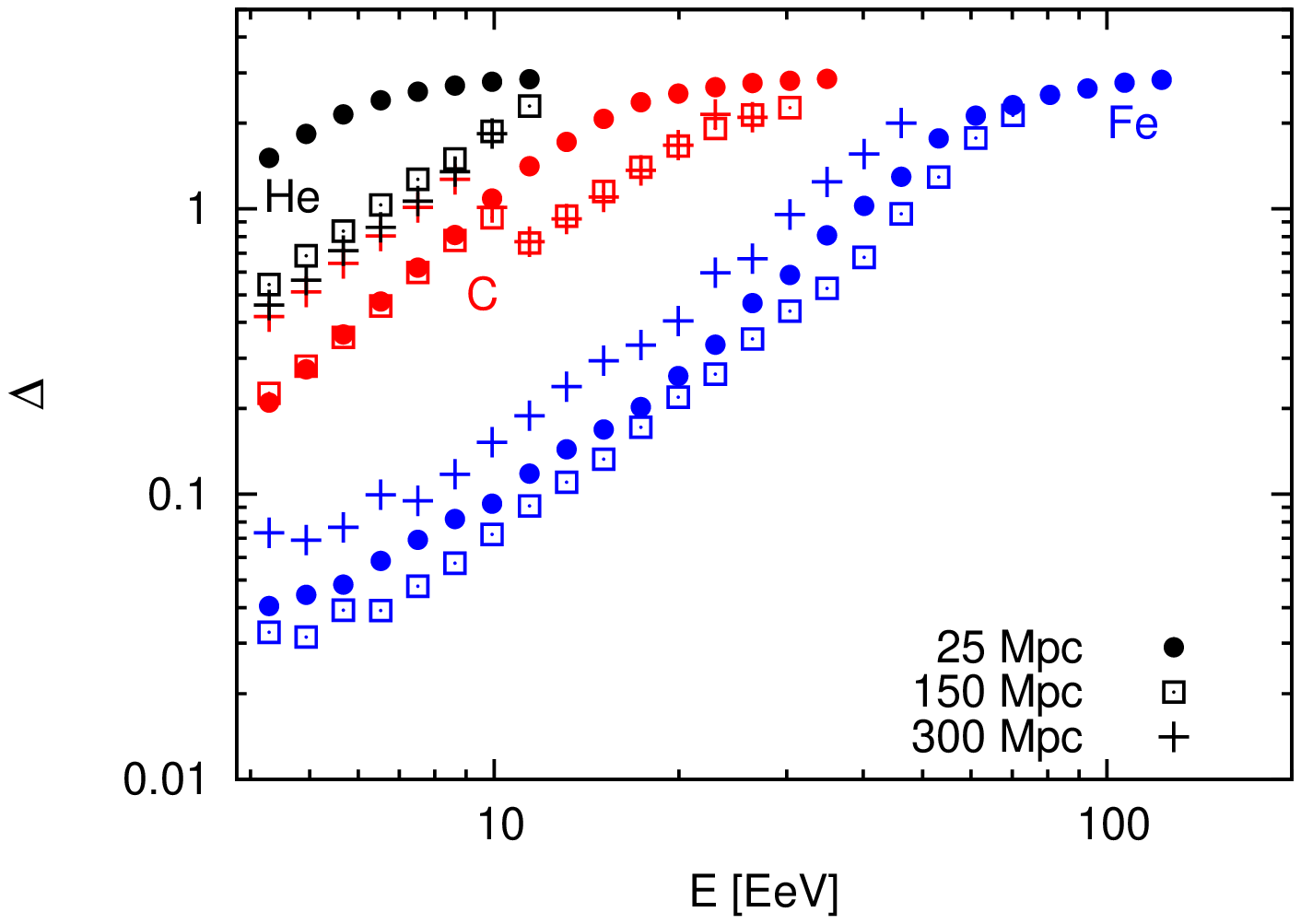}\epsfig{width=2.2in,angle=-90,file=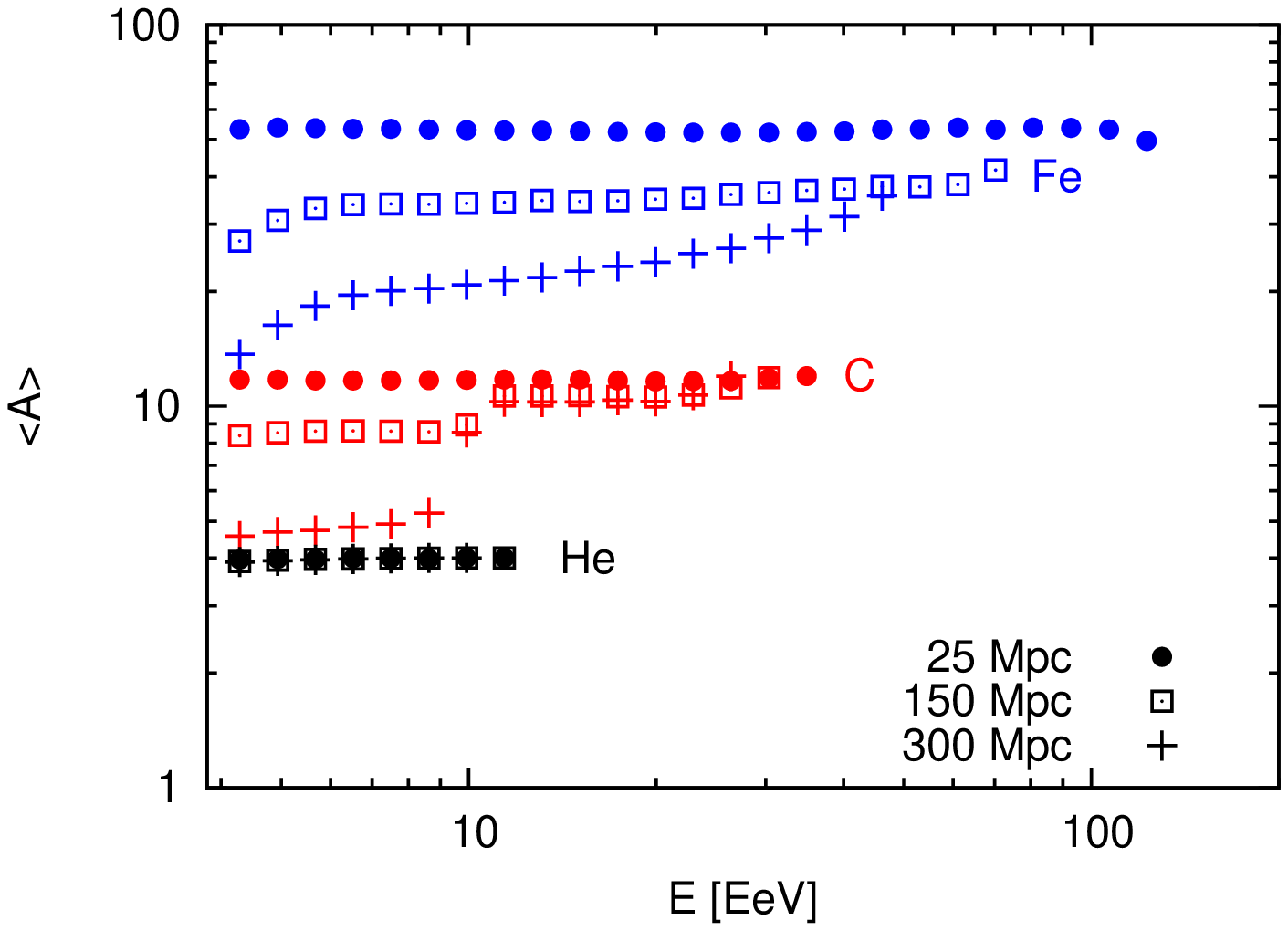}}
\vskip 1.0 truecm
\caption{Left: dipole amplitude as a function of the arrival energy for a source at distance of 25~Mpc (circles), 150~Mpc (squares) and 300~Mpc (plus signs) emitting He,  C or Fe nuclei. Right: mean mass as a function of the final energy for the same cases. A magnetic field strength $B=1$~nG is adopted.} 
\label{cfe}
\end{figure}

For example, we show  in the left panel of Figure \ref{cfe} the dipole amplitude for injected He, C and Fe nuclei coming from sources at three different distances, 25 Mpc, 150 Mpc and 300 Mpc, as a function of the arrival energy. 
Comparing the different curves  we see that for a given energy the lighter
 particles  lead to larger anisotropies. This arises because having larger rigidities they  suffer smaller deflections  than the heavier ones. For the closer distance displayed of 25~Mpc, for which the energy losses are less relevant, the anisotropies of the different components are quite similar for the same rigidity. For instance, shifting the Fe curve to the left by rescaling the energy by a factor 6/26 would closely match the results obtained for C.

 When considering sources farther away  the anisotropies first diminish with increasing distance, being smaller e.g. for 150~Mpc than for 25~Mpc, since the diffusive flux becomes more isotropic farther away from the source,  but then the anisotropy may increase again (see for instance the 300~Mpc curve for Fe).
This is the consequence of two effects. On one side the magnetic horizon\footnote{The magnetic horizon is essentially the distance from a source that a CR can travel diffusively in a time comparable to the age of the Universe. Hence, the density of CRs produced by the source should be strongly suppressed beyond the (energy dependent) magnetic horizon distance.}   enhances the gradient of the CR density at very large distances, and on the other hand, due to photo-disintegrations the contribution from lighter particles becomes significant for faraway sources. Since the photo-disintegrations reduce the energy of the leading fragment (by reducing its mass) but they leave essentially unchanged the particle rigidity (since $A/Z\simeq 2$ for nuclei heavier than H), this implies that at a given energy the average rigidity increases farther from the source, and this can lead to an increase in the resulting anisotropies with distance. In particular, 
a peculiar feature that is observed for intermediate nuclei, as in the C case, is a kink in the dipole amplitude curves appearing at $\sim 10$~EeV. This can be understood by looking at the right panel of Figure \ref{cfe}, where the mean mass of the arriving particles is plotted for the same injected nuclei and source distances considered in the left panel. The sudden drop in the mean mass for the injected C case appearing for energies below $\sim 10$~EeV is due to the effects of the Be disintegration to two alpha particles plus a nucleon. Since  the maximum C energy considered  is 36 EeV,  in this case the maximum energy for the alpha particle resulting from the photo-dissociation process would be about 12~EeV, and this is why the average mass number has a significant drop below this energy. Since the alpha particles appearing below 12~EeV have a relatively high rigidity, their contribution to the anisotropies is enhanced, explaining the rise in anisotropy observed below 12~EeV, and we also see that in this energy range the anisotropy from the faraway source would be comparable to that obtained in the case of injected He nuclei.

\begin{figure}[t]
\centerline{\epsfig{width=2.2in,angle=-90,file=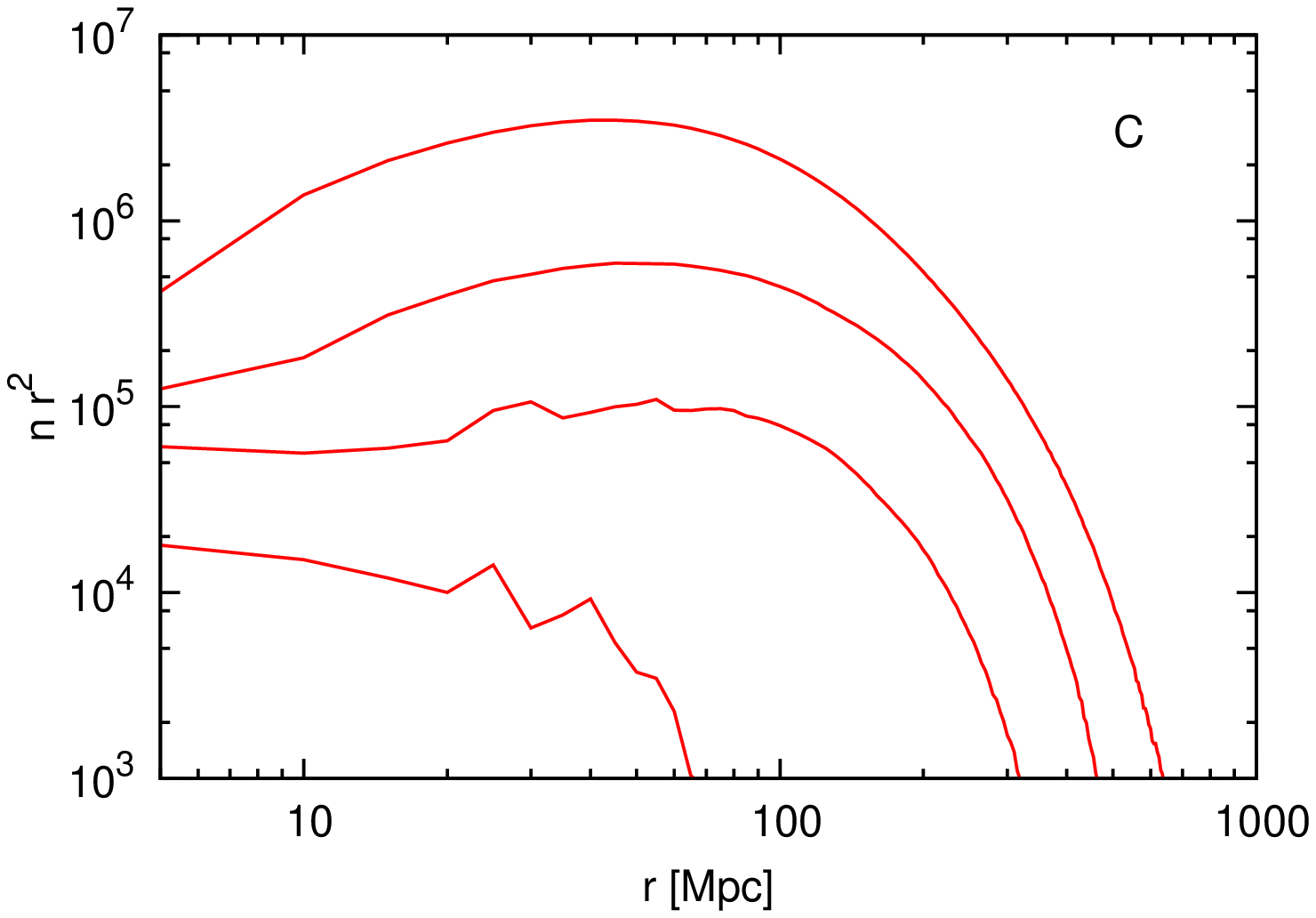}\epsfig{width=2.2in,angle=-90,file=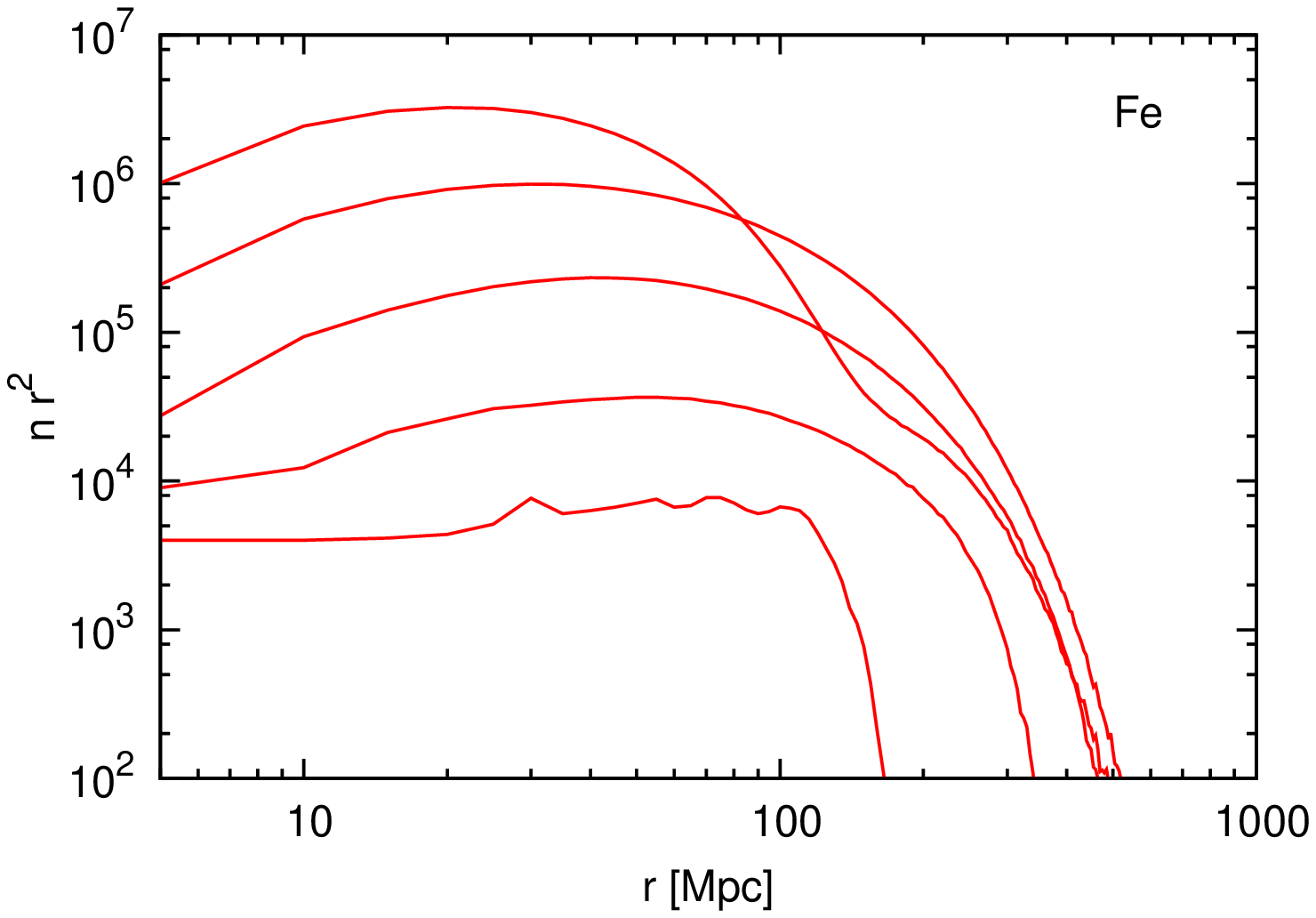}}
\vskip 1.0 truecm
\caption{Density of particles times $r^2$ (in arbitrary units) as a function of the distance to the source. The left panel is for C injected nuclei for energies of 4, 8, 17 and 34 EeV (from top to bottom). The right panel is for injected Fe nuclei and for energies of 4, 8, 17, 34 and 70 EeV.} 
\label{nr2}
\end{figure}

From the results of the simulations it is also possible to obtain the density of particles as a function of the distance to the source for different values of the energy. We show in Figure \ref{nr2} the results for injection of  C and Fe nuclei. At the highest energies, the density decreases as $n \propto r^{-2}$, as expected from quasi-rectilinear propagation, but particles quickly loose energy and cannot arrive from far away sources (farther than $\sim 50$~Mpc for 34~EeV carbons or $\sim 150$~Mpc for 70~EeV irons). For smaller energies
the losses are less important and particles can arrive from sources farther away. Moreover, an enhancement of the density with respect to the $r^{-2}$ law is evident at short distances, characteristic of the diffusive propagation, followed then by a steep decrease at large distances due to the magnetic horizon effect.

\section{Large scale anisotropies from a distribution of sources}

In the previous section the dipolar anisotropy from individual sources in the presence of a turbulent magnetic field was computed as a function of the source distance, type of nucleus injected and of the CR energy. In a realistic situation the total cosmic ray flux will probably originate from a set of several (or many) sources, and probably accelerating a mixture of different nuclei. The total dipolar component of the flux will mainly depend on the location and intensities of the nearest sources and on whether there is an inhomogeneous distribution of the sources at large scales. If there are several sources contributing to the flux, the dipolar anisotropy can be obtained from the superposition of the individual source dipoles.   

We first consider in this Section  the case in which just one type of nucleus, $j$, is injected by all the sources. The corresponding dipolar anisotropy is
\begin{equation}
{\vec \Delta}^{(j)} (E)=\sum_{i=1}^N \frac{n_i^{(j)}}{n_t}(E) {\vec \Delta}^{(j)}_{i} (E),
\label{diptot}
\end{equation}
where $N$ is the number of sources giving a non-negligible contribution to the flux at energy $E$, $n_i^{(j)}/n_t$ measures the fraction of the total flux coming from the $i$-th source and ${\vec \Delta}^{(j)}_i (E)$ is the dipole anisotropy of the flux from source $i$ as computed in the previous section for a given nucleus $j$.

\begin{figure}[t]
\centerline{\epsfig{width=2.in,angle=270,file=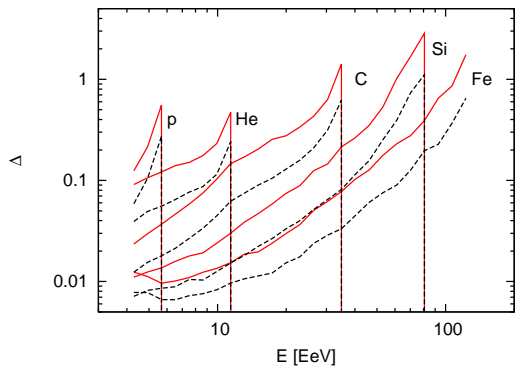}}
\vskip 1.0 truecm
\caption{Mean total dipole amplitude and its dispersion as a function of the energy for different injected nuclei and a density of sources $\rho = 10^{-5}$ Mpc$^{-3}$ (red solid lines) and $\rho = 10^{-4}$ Mpc$^{-3}$ (black dashed lines).}
\label{delt}
\end{figure}

In order to estimate the relative contribution to the anisotropy of the different sources as a function of their distance to the observer we will make the simplifying assumption that the sources are steady and have equal intrinsic intensities, so that for each energy the relative contribution to the flux from different sources  will only depend on the distance to the source $r_i$. 

The fact that the sources are distributed in different sky directions means that the vector sum in eq.~(\ref{diptot})  generally leads to a smaller dipole amplitude when many sources contribute. In the case that only few sources are relevant, the direction of these particular sources will determine the dipolar anisotropy, while if many sources are relevant, the overall large scale distribution of the sources, in particular whether the distribution has a non-vanishing dipole component, can have a significant effect. 

In order to quantify the total amplitude of the dipolar anisotropy we perform some simple simulations considering first  a homogeneous distribution of sources. Starting with one source at a random direction in the sky, that represents the closest source, we subsequently add new sources in random directions and compute the new total dipolar anisotropy using eq.~(\ref{diptot}). The radial distances from the observer to the sources are taken as the mean expected value for the $i$-th closest source in an homogeneous distribution, that is given by $\langle r_i \rangle = (3/4\pi\rho)^{1/3} \Gamma(i+1/3)/(i-1)!$, where $\rho$ is the density of sources. We show in Figure \ref{delt} the mean amplitude of the dipole obtained in 1000 simulations for two different values of the source density: $\rho = 10^{-5}$ Mpc$^{-3}$, for which the closest source is at a mean distance $\langle r_1 \rangle \simeq 25$ Mpc (solid lines), and  $\rho = 10^{-4}$ Mpc$^{-3}$, for which  $\langle r_1 \rangle \simeq 11$ Mpc (dashed lines)
and for the five nuclear species considered. For a larger density of sources the flux from each source is smaller, leading to smaller anisotropies.

\section{Dipolar anisotropy in a mixed-composition scenario}

We can now consider a more realistic scenario in which a distribution of sources,
that we again assume to have equal intensity and to be homogeneously distributed, accelerates cosmic rays with a mixed composition. For simplicity we will consider the same composition for all the sources, that we parameterize by the fraction $f_j$ of the differential energy flux contributed by nucleus $j$ at low energies (below the proton spectrum cutoff) of each of the five representative nuclei that we considered in the previous section. The expected dipolar anisotropy is then obtained by a weighted superposition of the amplitudes corresponding to each nuclei,  
\begin{equation}
{\vec \Delta} (E)=\sum_{i,j} f_j\frac{ n_i^{(j)}}{n_t}{\vec \Delta}^{(j)}_i (E),
\end{equation}
where  $n_t(E)=\sum_{i,j}f_j n_i^{(j)}(E)$ and  ${\vec \Delta}^{(j)}_i (E)$ is the dipolar anisotropy of source $i$ assuming that only the individual nuclei $j$ contribute, as  obtained in the previous section. 

\begin{figure}[t]
\centerline{\epsfig{width=2.1in,angle=270,file=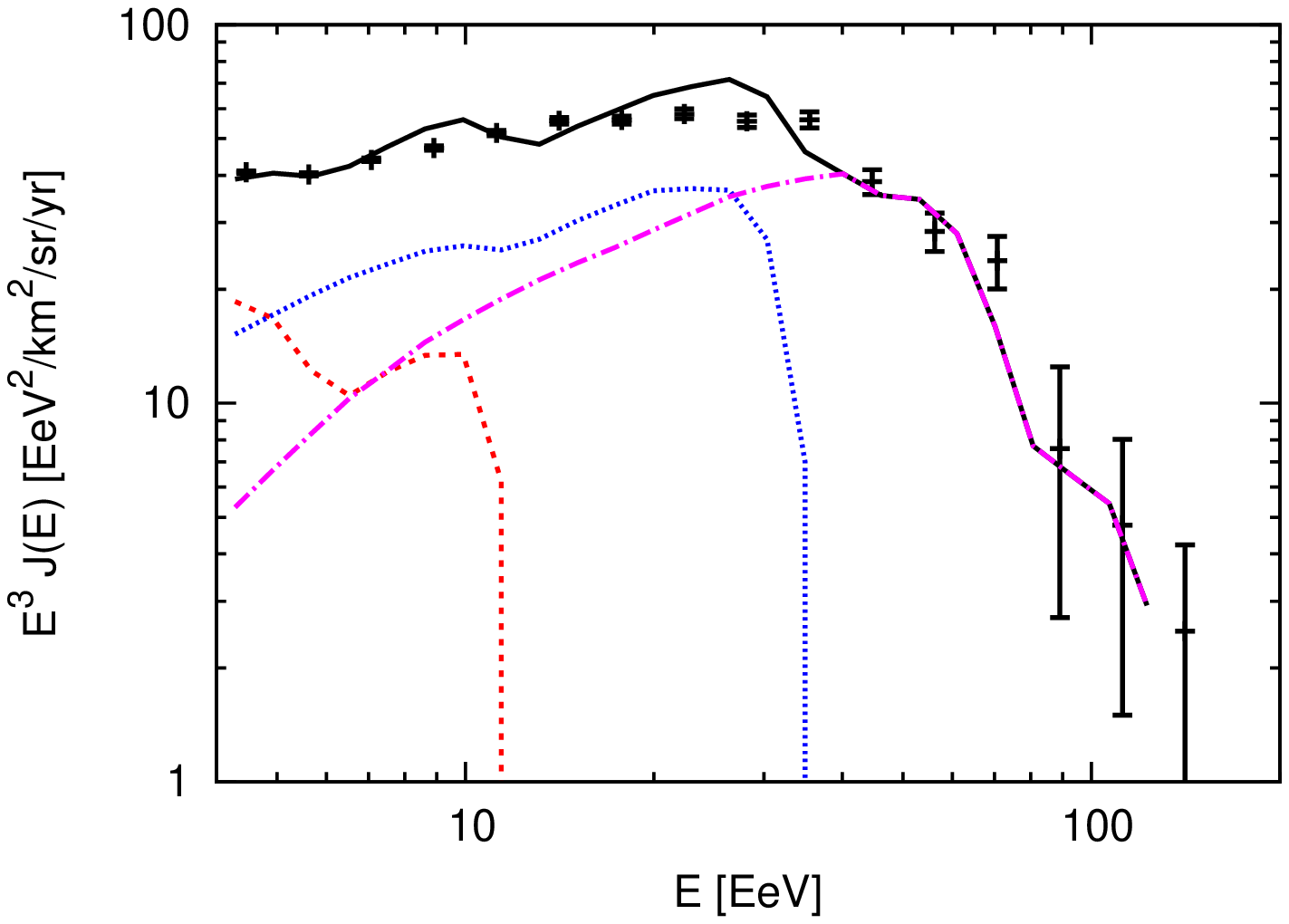},\epsfig{width=2.1in,angle=270,file=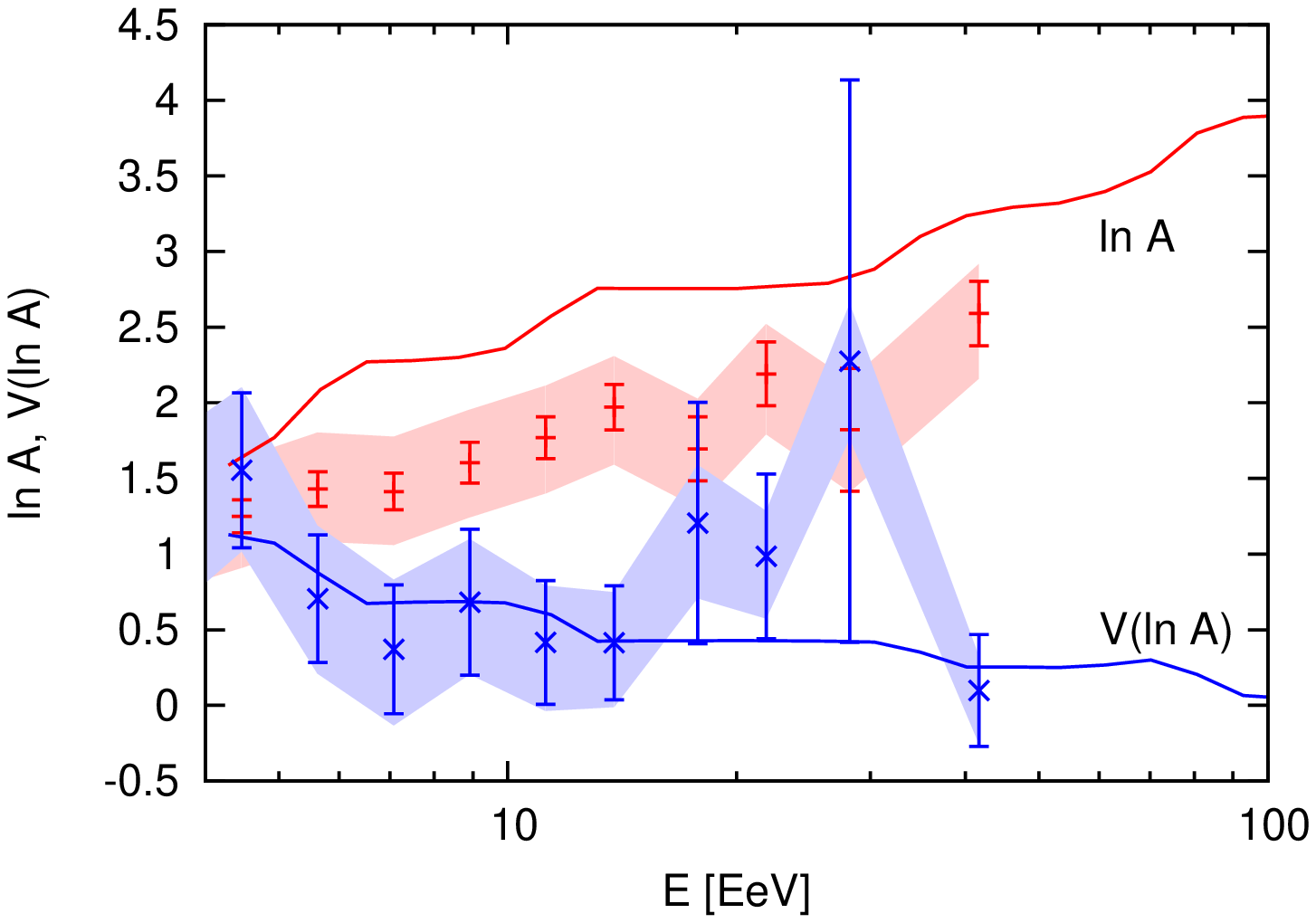}}
\vskip 1.0 truecm
\caption{Left: Spectrum of CRs. The data points are from the Pierre Auger Observatory \cite{augerspectrum}, the black solid line is for a homogeneous distribution of sources accelerating a mixed nuclei composition with $(f_p, f_{HE}, f_C, f_{Si}, f_{Fe}) = (0.19, 0.19, 0.4, 0.19, 0.03)$. The red dashed line is the contribution from nuclei with $A \le 4$, the blue dotted line is that from nuclei with $5 \le A \le 16$ and the purple dot-dashed line is for $17 \le A \le 56$. Right: composition at Earth in terms of $\langle \text{ln}A\rangle$ and its variance, $V(\text{ln}A)$. The data points are from the Pierre Auger Observatory \cite{augercomposition1} based on an interpretation of air-shower data using EPOS-LHC. Error bars correspond to the statistical uncertainties and color bands to the systematic ones. The solid lines are the predictions for the same model used in the left panel. }
\label{speclna}
\end{figure}

As an example we will show the results for a particular set of nuclei fractions at the sources that lead to a cosmic ray spectrum similar to that measured by the Pierre Auger Observatory \cite{augerspectrum}. The left panel of Figure \ref{speclna} shows the overall spectrum of particles reaching the observer from sources accelerating particles with a spectrum $E^{-2}$ up to maximum energy $E_{max}= 6Z$~EeV and injected fractions $f_p=f_{He}=f_{Si}=0.19, f_C=0.4$ and $f_{Fe}= 0.03$,  propagating through a homogeneous turbulent magnetic field with Kolmogorov spectrum, coherence  length of 1 Mpc and root mean square amplitude of 1~nG. Given the assumed  $E^{-2}$ spectrum and the sharp cutoffs adopted the spectrum obtained with the chosen fractions is indeed quite acceptable. Also shown in  Figure \ref{speclna} are the separate contribution to the spectrum from the three ranges of masses, corresponding to   $A \le 4$, $5 \le A \le 16$ and $17 \le A \le 56$.

The composition as a function of the energy also follows the general trends measured by the Pierre Auger Observatory, as it can be seen in the right panel of Figure \ref{speclna}, where the mean $\langle \ln A \rangle$ and variance $V(\ln A)$ are shown for the model and are compared to the values obtained from the  data adopting the EPOS-LHC hadronic interaction model.

\begin{figure}[t]
\centerline{\epsfig{width=2.1in,angle=270,file=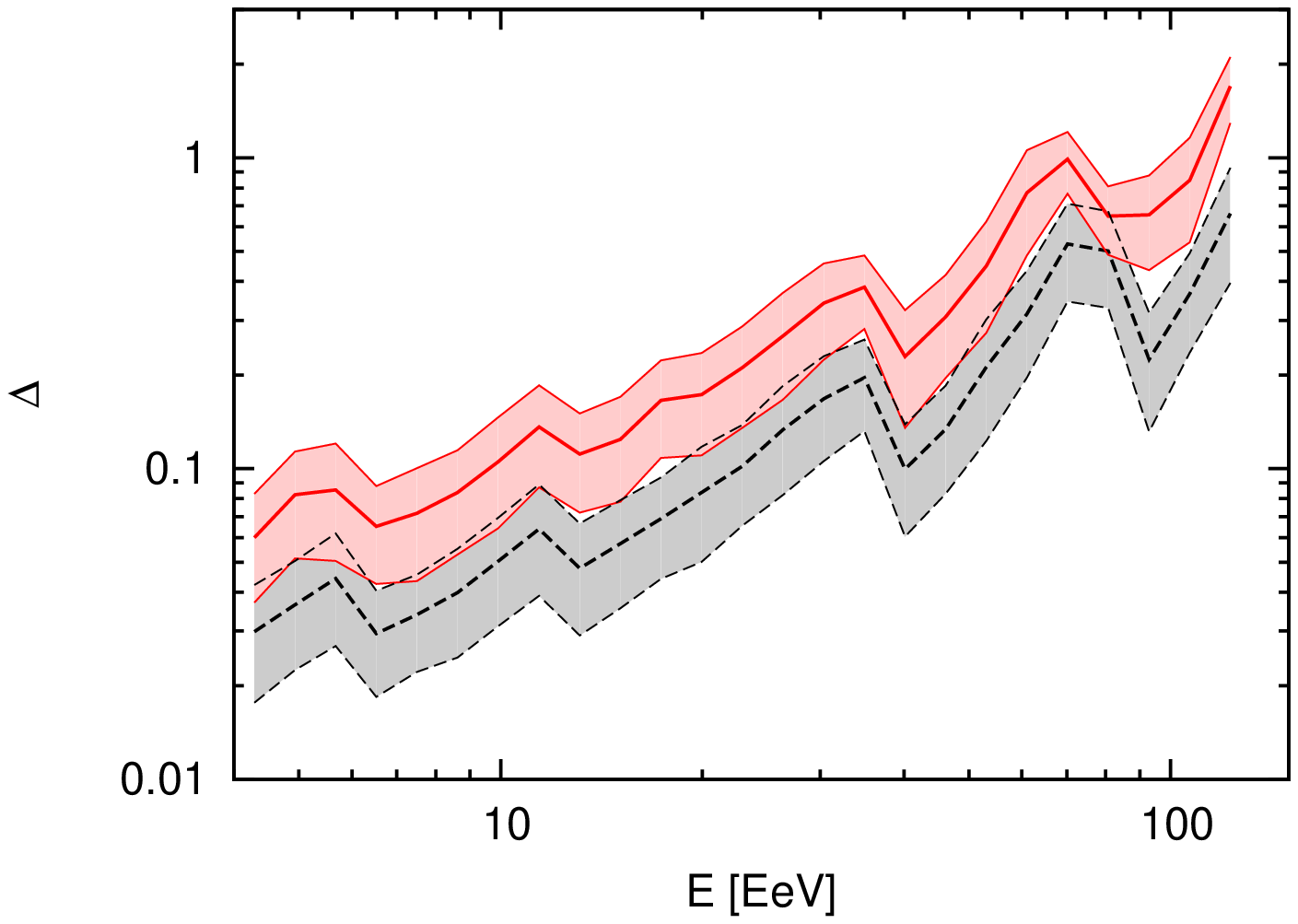},\epsfig{width=2.1in,angle=270,file=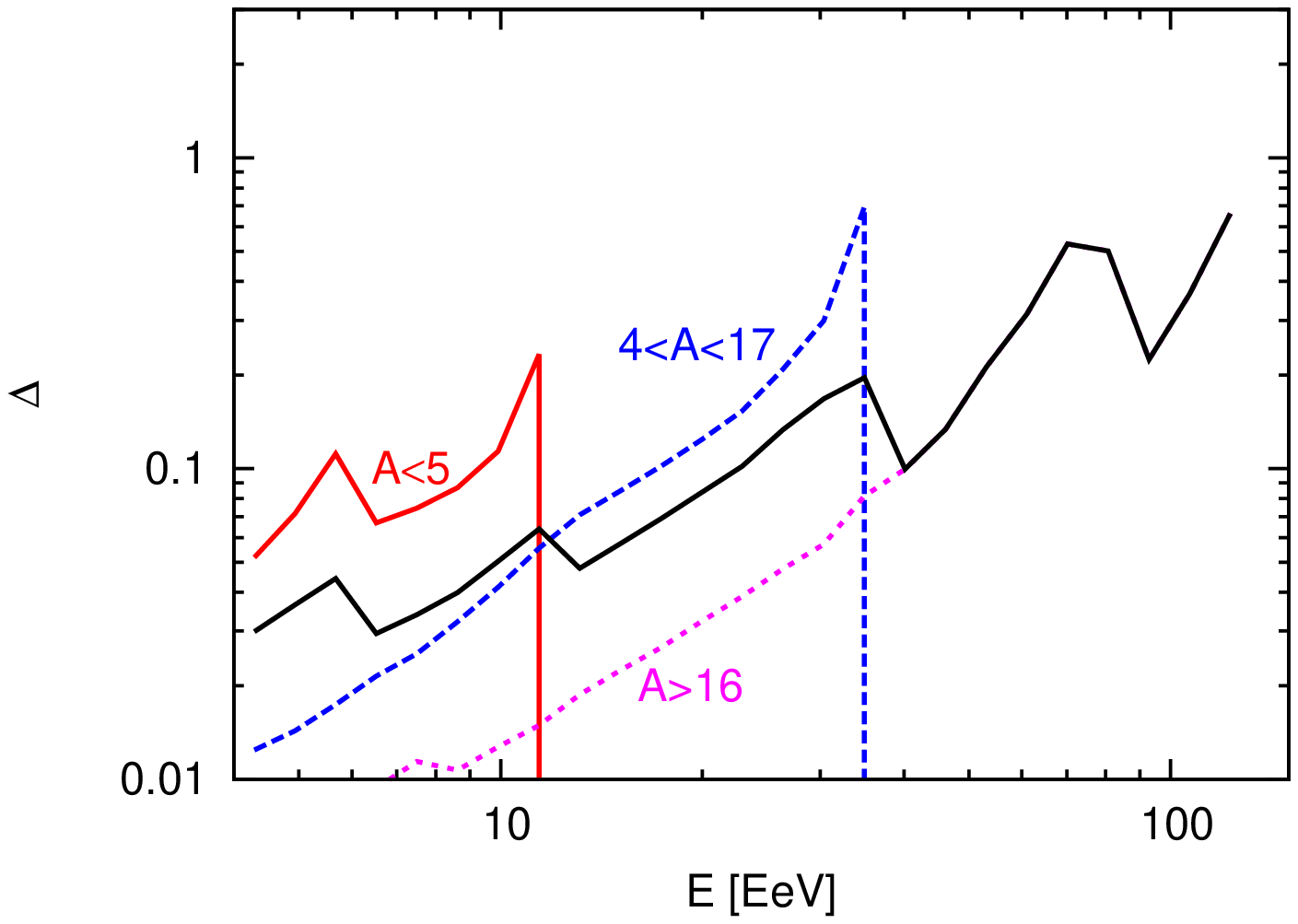}}
\vskip 1.0 truecm
\caption{Left: Mean total dipole amplitude and its dispersion as a function of the energy for a homogeneous distribution of sources with density $\rho = 10^{-5}$ Mpc$^{-3}$ (red solid lines) and $\rho = 10^{-4}$ Mpc$^{-3}$ (black dashed lines) accelerating a mixed nuclei composition with $(f_p, f_{HE}, f_C, f_{Si}, f_{Fe}) = (0.19, 0.19, 0.4, 0.19, 0.03)$. Right: mean dipole amplitude discriminating by composition at Earth, in the mass ranges as indicated by the labels, for the case of sources with density $\rho = 10^{-4}$ Mpc$^{-3}$.}
\label{deltmixed}
\end{figure}

The total dipolar anisotropy from homogeneously distributed sources with spatial density  $\rho = 10^{-5}$ Mpc$^{-3}$ and $\rho = 10^{-4}$ Mpc$^{-3}$ and accelerating the nuclei with the above mentioned fractions is shown in the left panel of Figure \ref{deltmixed} as a function of the energy. Here the  mean and dispersion of the dipolar amplitude  obtained from 500 simulations are plotted. The sharp features appearing in the curves at the energies corresponding to maximum acceleration for the five elements considered are an artifact of the sharp cutoffs adopted, and are expected to be smoothed out in scenarios with softer suppressions, such as exponential ones.

In the right panel of Figure \ref{deltmixed} we plot the dipole amplitude separately for different mass groups. We divide the composition observed at Earth in three sets,  $A \le 4$, $5 \le A \le 16$ and $17 \le A \le 56$, the same ranges used to illustrate the different contributions to the spectrum in the left panel of Figure \ref{speclna}. We show the dipolar anisotropy amplitude for each mass group as a function of the energy for the case of a homogeneous distribution of sources with density $\rho = 10^{-4}$ Mpc$^{-3}$. The amplitude in the lighter components is enhanced relative to the total one since the heavier fraction is more isotropically distributed at lower energies.

\begin{figure}[t]
\centerline{\epsfig{width=2.5in,angle=270,file=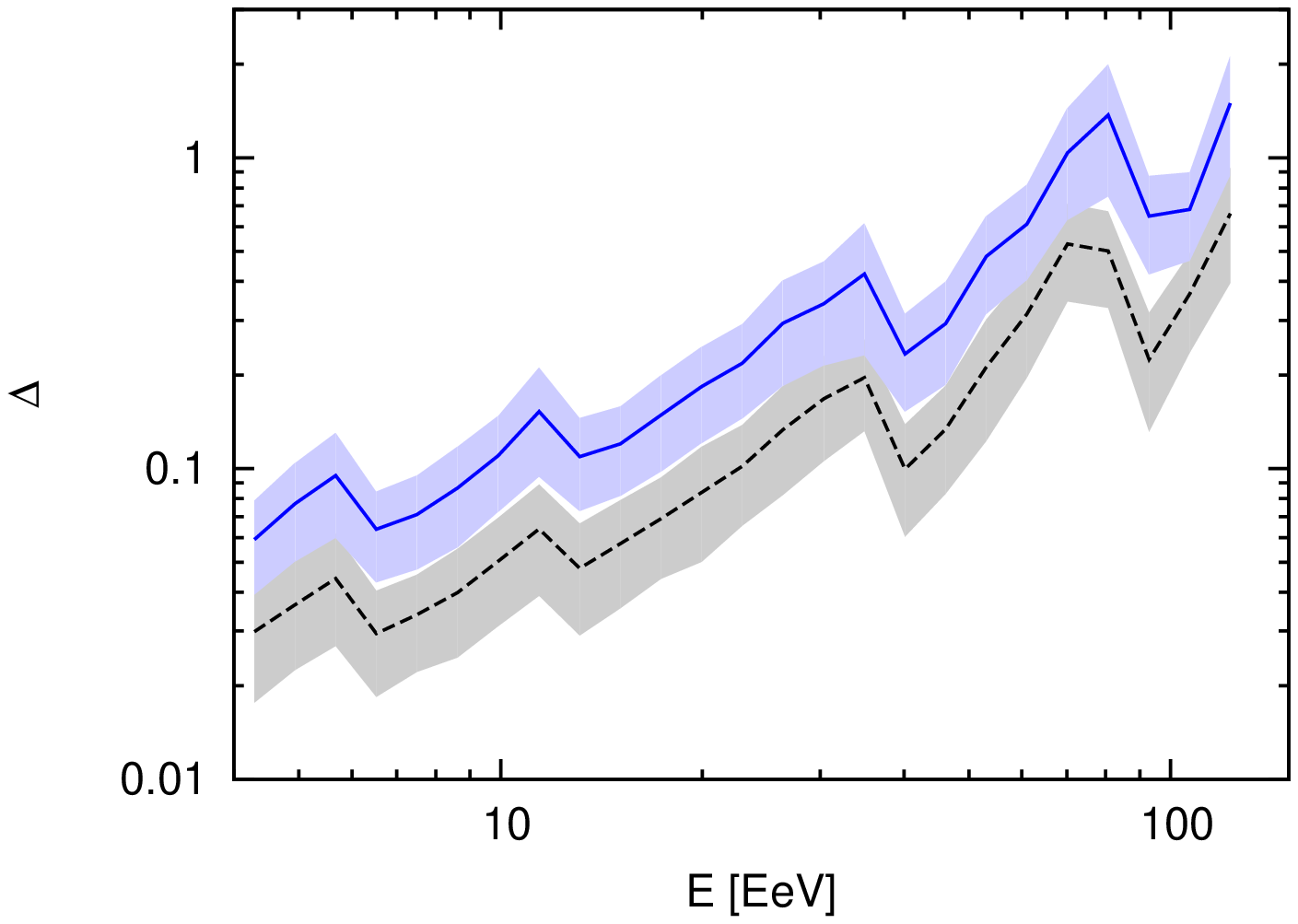}}
\vskip 1.0 truecm
\caption{Mean total dipole amplitude and its dispersion as a function of the energy for a distribution of sources with density $\rho = 10^{-4}$ Mpc$^{-3}$ homogeneously distributed in space (black dashed line) and following the local matter distribution (blue solid line) that accelerate a mixed nuclei composition with the same injected fractions as in Figure \ref{deltmixed}.}
\label{dip2mrs}
\end{figure}

The previous results hold for the case of  homogeneously distributed sources. In the case that the sources themselves have an inhomogeneous distribution around the observer, in particular if their distribution has a non-vanishing dipole, a further contribution to the anisotropy is expected. For instance, if the distribution of nearby cosmic ray sources follows the local distribution of matter a non-vanishing dipole is expected. The dipolar component of the matter distribution is indeed known to be responsible for the Local Group peculiar velocity with respect to the rest frame of the CMB, that actually gives rise to the observed CMB dipole. 
The dipolar component of the mass distribution in our neighborhood has been estimated using different catalogs of galaxies, as for example the 2 Micron All-Sky Redshift Survey (2MRS),  showing that the resulting dipole seemingly converges when sources up to a distance $\sim 90$ Mpc are included \cite{2mrs}.

The effect of the local inhomogeneity of the source distribution  in the predicted large scale anisotropies can be included in the simulations by choosing the positions of the sources in our neighborhood from some catalog representing the local distribution of matter.  To describe the local distribution of matter we use a volume limited subsample\footnote{Considering only objects with $d <$ 100 Mpc and absolute magnitude in the K band $M_K < -23.4$.} of the 2MRS catalog up to 100~Mpc \cite{hu12}. We have then selected the position of the required number of sources (according to the density considered) from this subsample of  2MRS galaxies. On the other hand, the locations of sources farther away were assumed to be  isotropically distributed. We show in Figure \ref{dip2mrs} the change in the mean dipole amplitude when the inhomogeneous source distribution is considered, adopting the same mixed-composition scenario analyzed above, a density  $\rho  = 10^{-4}$ Mpc$^{-3}$ and a turbulent field of $B= 1$ nG. Compared to the  homogeneous case, an enhancement of the dipole amplitude by a factor of about 2 is observed.

\section{Anisotropies and the strength of the magnetic field}

In the previous sections we computed the resulting anisotropies, both in the case of a single source as well as  for a superposition of sources, adopting a reference value of 1\,nG for the RMS turbulent extragalactic magnetic field $B$.
 We now discuss how the predictions would be affected for different assumed strengths of the magnetic field.

\begin{figure}[t]
\centerline{\epsfig{width=2.1in,angle=270,file=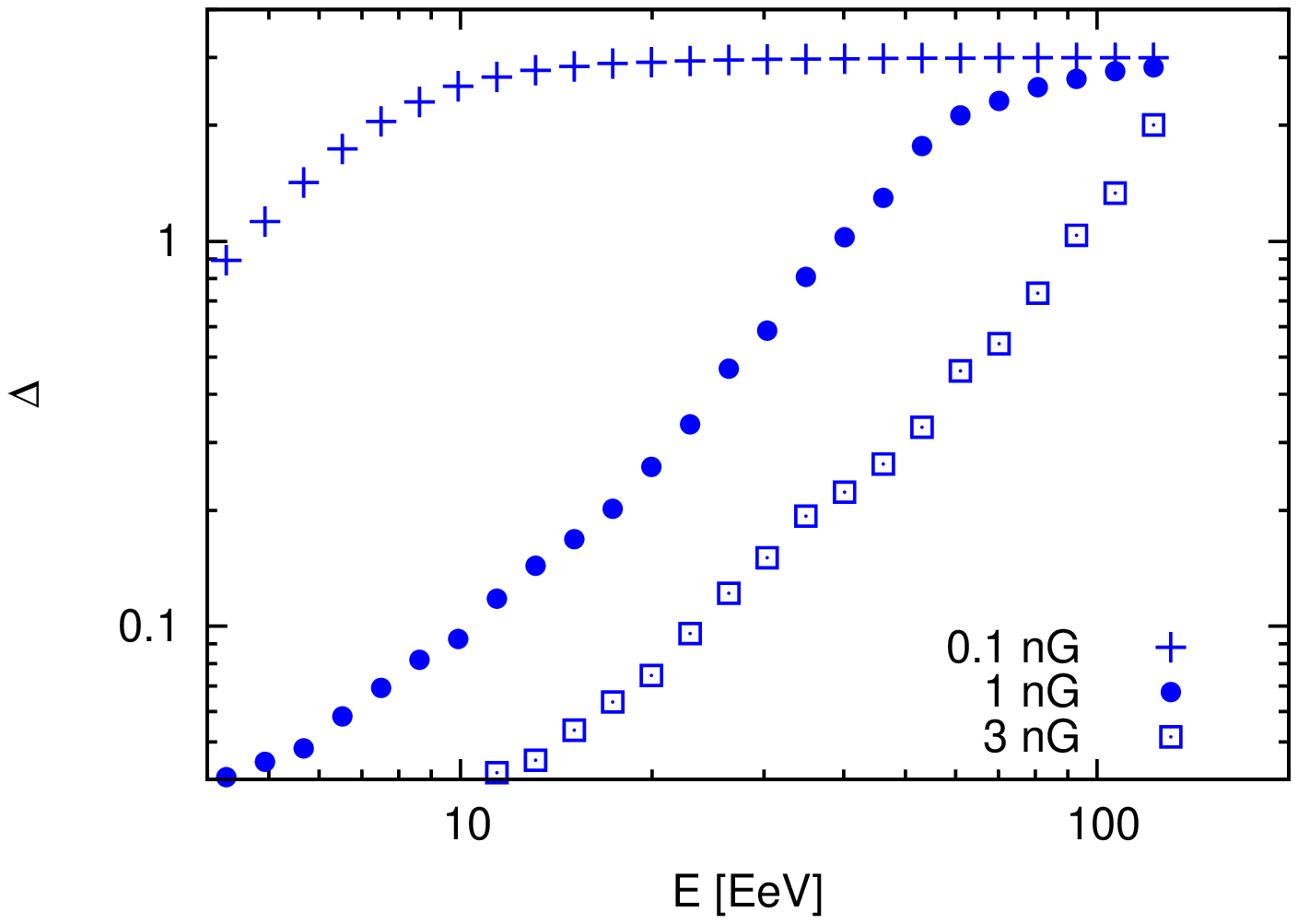},\epsfig{width=2.1in,angle=270,file=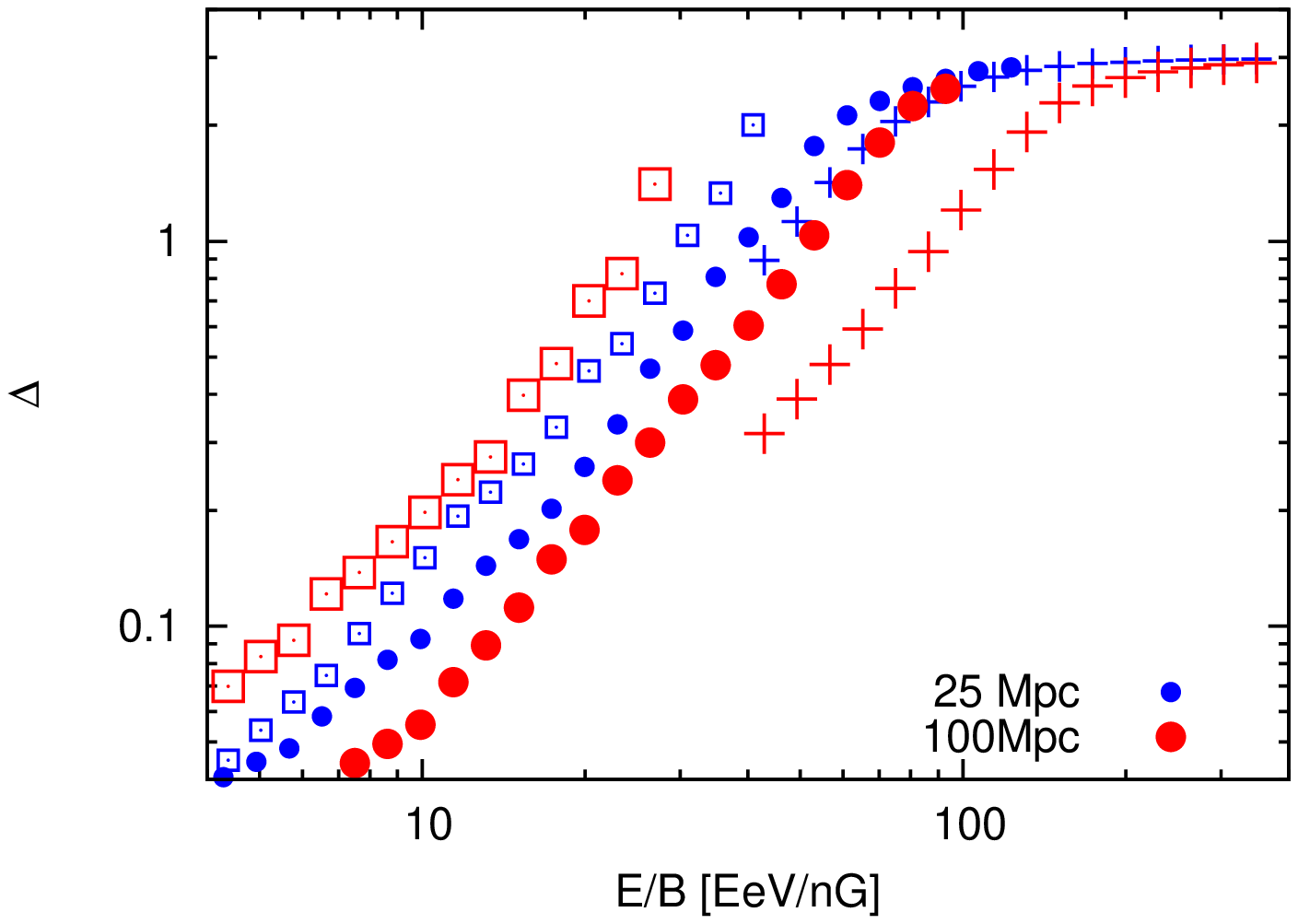}}
\vskip 1.0 truecm
\caption{Left: Dipolar anisotropy as a function of the arrival energy for an Fe source at 25 Mpc in the presence of a turbulent magnetic field of 0.1 nG, 1 nG and 3 nG.  Right: In blue (small symbols) the same results as in the left plot but displayed as a function of $E/B$. In red (large symbols) the corresponding results for a source at 100 Mpc.}
\label{bdep1}
\end{figure}

The left panel of fig. \ref{bdep1}  shows the energy dependence of the anisotropy of an individual source of Fe nuclei at  a distance of 25 Mpc, under the assumptions of $B=0.1$, 1 and 3\,nG.
As expected, the resulting anisotropy values  in the different cases are comparable if one rescales the energies by the inverse of the assumed magnetic  field strength. This results because the diffusion coefficient $D$ is just a function of $E/E_c$, and the critical energy is proportional to $B$.
To better appreciate this behavior we  show in the right panel the resulting anisotropies as a function of $E/B$ for sources at 25 and 100 Mpc. The approximate scaling is apparent, but is however violated  by the effects of interactions with the background radiation, since the interaction cross sections just depend on $E$. Also the fact that the surviving leading fragment can have a slightly different rigidity (since $A/Z$ may vary) can affect the final anisotropy.
We see then that the overall impact of photo-disintegrations is more pronounced for sources farther away and the anisotropies for larger magnetic fields are larger than what would be suggested by the simple scaling. This increase is mostly due to the fact that for stronger $B$ fields the distance travelled by the CRs from the source is larger, and hence the photo-disintegration effects are stronger, so that for a given arrival energy the actual rigidity of the arriving particles will be larger, and hence the anisotropy will be stronger.

Turning now to the case of a superposition of sources, here an interesting effect appears that leads to a reduced dependence of the dipolar amplitude with respect to the strength of the magnetic field. Indeed, for a given nearby source (closer than the magnetic horizon), as the value of  $B$  is increased  the individual contribution of that source to the CR density increases because its  density is enhanced by the diffusion while that from far away sources is suppressed. On the other hand, the value of the dipolar component of its anisotropy decreases in such a way that both changes compensate each other to a large extent. In particular, in the diffusion regime and neglecting interaction and redshift effects, one has that the CR density from one source satisfies
\begin{equation}
n_i= \frac{Q}{4\pi r_i D},
\end{equation}
with $D$ the diffusion coefficient, while its contribution to the dipole is
\begin{equation}
\Delta_i= \frac{3 D}{r_i}.
\end{equation}
We see then that for a superposition of sources
\begin{equation}
{\vec \Delta}=\sum_i\frac{n_i{\vec \Delta_i}}{n_{tot}}\simeq \frac{1}{n_{tot}}\sum_i\frac{3 Q \hat r_i}{4\pi r_i^2},
\label{aniconst}
\end{equation}
which is independent of the value of $D$, and thus of the magnitude of $B$, and the result is  similar to what would be obtained in the absence of magnetic fields. We have here used that, due to the propagation theorem \cite{al04}, the total contribution to the CR density $n_{tot}$ is independent of the value of $B$, which is valid as long as the source density is uniform and the inter-source distance is smaller than the diffusion and interaction lengths.
We have  checked numerically that the anisotropy result in eq. (\ref{aniconst}) also holds, as long as interaction and redshift effects are ignored, in the regime of quasi-rectilinear propagation.

\begin{figure}[H]
\centerline{\epsfig{width=2.5in,angle=270,file=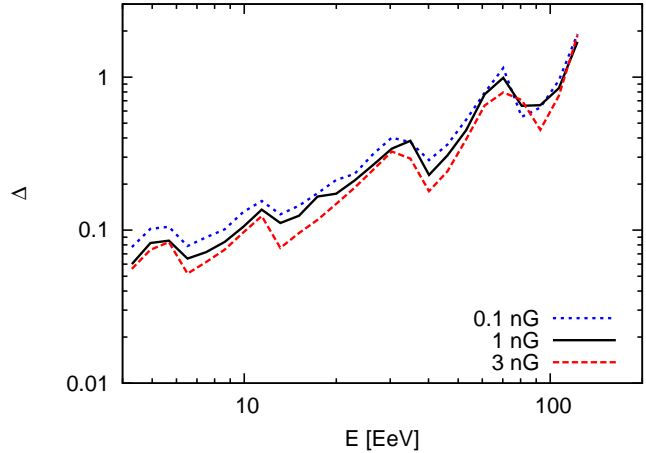}}
\vskip 1.0 truecm
\caption{Mean total dipole amplitude as a function of the energy for a homogeneous distribution of sources with density $\rho = 10^{-5}$ Mpc$^{-3}$  accelerating a mixed nuclei composition with $(f_p, f_{HE}, f_C, f_{Si}, f_{Fe}) = (0.19, 0.19, 0.4, 0.19, 0.03)$ in the presence of a turbulent magnetic field of 0.1 nG, 1 nG and 3 nG.}
\label{bdep3}
\end{figure}

In fig. \ref{bdep3} we show the resulting anisotropies, including interaction losses and redshift effects, for the case of a uniform source distribution with density $10^{-5}$ Mpc$^{-3}$, for the same composition mixture considered in fig. \ref{speclna} and for the three values $B=0.1$, 1 and 3\,nG. We see that differences in the dipolar component of the anisotropy by up to a factor of 2  can arise due to the interaction effects.

It is important to note however that when the deflections from the nearby sources become small, i.e. at high rigidities, the CRs arriving from those sources will be quite localized in the sky. Hence, although the dipolar component of the arrival direction distribution may have only a moderate dependence on the actual strength of the $B$ field, the detailed distribution will be quite different in both cases. In particular, higher harmonics will become more sizeable for weaker $B$ fields and the individual contributions from the nearby sources would become distinguishable in this case. 

Finally, let us mention that the additional deflections of the CRs due to the galactic magnetic field, not included in this work, would modify the CR dipole amplitude and direction as well as generate higher order multipoles in the arrival directions distribution \cite{ha10}, but for the energies considered here the induced change in the dipole amplitude is expected to be small. 

\section{Conclusions}

The low level of anisotropies in the distribution of arrival directions of UHECRs \cite{Augerdensitybounds,AugerAnisoHE,TAanisoHE} challenges scenarios with a relatively small density of 
cosmic ray sources and a light composition. In addition, there are hints of a dipole anisotropy with amplitude around 7\% for energies above 8 EeV \cite{AugerDipole}. 

We have shown in paper I \cite{I}, where we assumed the CRs to be protons,  that a dipole anisotropy with amplitude between 3\% and 10\% at 10 EeV is expected in scenarios in which CRs that originate in extragalactic sources 
diffuse in turbulent magnetic fields. This result was derived for source densities between  $10^{-4}$ Mpc$^{-3}$ and  $10^{-5}$ Mpc$^{-3}$,  magnetic fields with strength of order 1 nG and coherence length of the order of 1 Mpc.  Here we extended the calculation of the predicted dipole anisotropy as a function of CR energy to the case of nuclei. We developed a method to evaluate the anisotropy across different regimes, from spatial diffusion to quasi-rectilinear propagation, accounting for energy losses and nuclei fragmentation. We applied the method to mixed-composition scenarios in which different cosmic ray nuclei are accelerated up to the same maximum rigidity. With sufficient magnetic turbulence, each component contributes significantly only in the energy range between its low-energy cut-off due to a magnetic horizon effect and its maximum acceleration limit. We illustrated the method in a scenario with maximum acceleration energy $E_{max}=6Z$~EeV, and fractions of different nuclei $f_p=f_{He}=f_{Si}=0.19, f_C=0.4$ and $f_{Fe}= 0.03$ injected with an $E^{-2}$ spectrum, which provides an acceptable match to the observed spectrum \cite{augerspectrum} and composition \cite{augercomposition1} derived from measurements by the Pierre Auger Observatory.

Naively one could expect a smaller dipole anisotropy at a given energy in a scenario with nuclei instead of just protons, due to the larger magnetic deflections. We found instead that the mixed-composition scenario considered here also leads to dipole amplitudes close to 5--10\% at 10 EeV, and larger values at higher energy, for the same source densities and magnetic fields as quoted above for the proton-only model, and showed that the dependence with the actual value of the magnetic field strength is moderate. As can be seen in Figure \ref{delt}, the dipole amplitude in the arrival directions for different injected nuclei are comparable for similar rigidities, even though energy loss and fragmentation processes introduce important differences. In the mixed-composition scenario considered here, the anisotropy due to a given type of nucleus below the respective $E_{max}$ is on one side enhanced  due to the suppression of faraway sources for energies close to $E_{max}$, but on the other hand the overall anisotropy is also reduced by the effects of heavier, more isotropic components, leading to the trend depicted in  Figure \ref{deltmixed}.
The inhomogeneity in the local source distribution enhances the predicted dipole amplitude by a factor of about 2 compared to the homogeneous case, as shown in Figure
\ref{dip2mrs}. Since this mixed-composition model predicts dipole amplitudes in the lighter components larger than the total one (at energies below the respective $E_{max}$), an observatory with capability for mass-discrimination could provide additional tests of this scenario and improve the determination of its parameters.

\section*{Acknowledgments}
 Work supported by CONICET and ANPCyT, Argentina. We thank the members of the Auger Collaboration for useful discussions.

\end{document}